\documentclass[arxiv,usenatbib]{iupartex}
\usepackage{longtable}

\usepackage{newtxtext,newtxmath}
\usepackage[T1]{fontenc}

\title[The Period Variation of KR Cyg]{Monitoring the Period Variation of KR Cyg Eclipsing Binary}

\author[Geldi et al.]{%
M. Geldi$^{1\cc}$,\orcid{0009-0004-0863-9957}
K. Çalışkan$^{1}$,\orcid{0009-0007-7857-6850}
and
E. Sipahi$^{1}$\orcid{0000-0003-1661-4907}
\affsep \\
$^1$ Ege University, Faculty of Science, Department of Astronomy and Space Sciences, 35100 Bornova, Izmir, Turkiye\\
}

\corres{M. Geldi}{mehmetgeldi03@gmail.com}

\pubyear{20XX}

\doiheader{XXXXXXX/PAR.20XX.00000}
\date{
	\pSubmit{20.04.2025} 
	\pRevReq{07.05.2025}
	\pAccept{00.00.0000}
	\pPubOnl{00.00.0000}
}
\volume{3}
\volnumber{1}

\begin{document}
\label{firstpage}
\pagerange{\pageref*{firstpage}--\pageref*{lastpage}}
\maketitle

\begin{abstract}
This study examines the period variation of the eclipsing binary system KR Cyg, a near-contact binary characterized by its short orbital period and significant stellar interactions. The precise orbital parameters of the system have been determined through light curve data collected periodically since 1999 and were published in our previous studies. Additionally, minimum light observations have been continuously monitored to analyze the period variation of the system. In our earlier work, the masses of the primary and secondary components were determined as $2.88\pm0.20 M_{\odot}$ and $1.26\pm 0.07 M_{\odot}$, with corresponding radii of $2.59\pm 0.06 R_{\odot}$ and $1.80\pm 0.04 R_{\odot}$. The bolometric albedo and effective temperature of the less massive star were also investigated, and deviations caused by mutual illumination effects were identified. These results provide valuable insights into the evolutionary status of KR Cyg and shed light on the dynamic processes that occur in near-contact binary systems. Using updated eclipse timings, an (O-C) diagram was constructed, revealing both long-term trends and periodic oscillations. These cyclical variations are thought to indicate the presence of a tertiary component affecting the system via the light-time effect. Additional observations and modeling are recommended to confirm the tertiary hypothesis and further refine the parameters of the system.
\end{abstract}




\section{Introduction}
Near-contact binary systems (NCBs) represent a fascinating class of stellar binaries where at least one component is close to filling its Roche lobe, often leading to significant tidal interactions and mutual distortion. These systems are typically characterized by short orbital periods and exhibit complex light curve variations due to effects such as ellipsoidal modulation, reflection, and sometimes partial eclipses. NCBs serve as crucial laboratories for studying stellar evolution, mass transfer processes, and the influence of proximity on stellar structure and behavior. Their diverse configurations often position them as precursors to more evolved systems, such as contact binaries or cataclysmic variables. Understanding the physical and dynamical properties of NCBs provides valuable insights into binary evolution scenarios and their role in broader astrophysical contexts.

KR Cyg is an EB-type eclipsing binary system initially identified by \citet{schneller} in the early 1930s. Visual light curves were later documented by \citet{lasso}, \citet{gapo1953}, and \citet{tse1954}. Photographic light curves were produced by \citet{watchmann} and \citet{nekra}, whose studies indicated that KR Cyg is an Algol-type binary. Subsequent observations by \citet{vetesnik} introduced the first photoelectric light curves for the system, determining an orbital period of 0.8451538 days. Based on these light curves, Vetešnik categorized the system as $\beta$ Lyrae-type, comprising a B9-type primary star and an F5-type secondary star. However, \citet{koch} assigned the secondary star a spectral type of G4, while \citet{hill} conducted the first spectroscopic analysis, suggesting the system is closer to type A3. In 1980, \citet{wilson} analyzed Vetešnik’s light curves and calculated the first photometric solution, yielding a mass ratio of $q=0.478$. A statistical analysis by \citet{giu1983} grouped KR Cyg among ‘a’ class Algol systems, which includes classical semi-detached binaries and Algols with subgiant secondaries. \citet{naim} applied Fourier analysis to derive the geometric and physical properties of the system, reporting a slightly adjusted photometric mass ratio of $q=0.51$. Later, \citet{khopolov} classified the system as A2V. In 2004, \citet{sipahi} presented three-colour photometric data, further refining the system's parameters. Their WD-based analysis revealed a photometric mass ratio of $q=0.43$, suggesting that the cooler, less massive secondary nearly fills its Roche lobe. \citet{shaw} classified KR Cyg as a near-contact binary (NCB), a group where both stars either fill or nearly fill their Roche lobes. At a later time, \citet{sipahii} compiled all known light minimum times and analysed the system's orbital period. Her findings revealed periodic oscillations in the (O-C) data, with an amplitude of 0.001 days and a cycle lasting approximately 76 years, which she attributed to a hypothetical third component. Shortly after \citet{sipahiii} presented multi-colour, five-year photometric light curves and radial velocity measurements for the near-contact binary system KR Cyg. 

In this study, the component masses were determined as $2.88\pm 0.20 M_{\odot}$ and $1.26\pm 0.07 M_{\odot}$, with corresponding radii of $2.59\pm 0.06 R_{\odot}$ and $1.80\pm 0.04 R_{\odot}$. Additionally, they investigated the empirical determination of the albedo and effective temperature of the cooler, less massive star in KR Cyg, alongside a comparison with two similar near-contact binaries, AK CMi and DO Cas. Discrepancies between the observed and computed fluxes are attributed to mutual illumination effects, where the heated surface layers of the illuminated star experience variations in bolometric albedo, limb-darkening coefficients, and gravity-brightening exponents. Interestingly, the derived effective albedos are generally lower than those predicted for stars with convective envelopes. It is important to note that these findings are preliminary and warrant further investigation. \citet{tvar} have also published a study on the period variation of the system. In this study, the light curve variations and period changes of the KR Cyg eclipsing binary system were analysed using the photometric observations from 2024 and the updated minimum time. This study aims to investigate the presence of the third component in the system through (O-C) analysis using approximately 95 years of minimum time data.

\section{Observations}
Observations were acquired with a thermoelectrically cooled ALTA U +47 1024 × 1024 pixel CCD camera attached to a 40 cm Schmidt-Cassegrains type MEADE telescope at Ege University Observatory. The observations made in the $V$ band were continued on September 27, 28, and 29, in the season of 2024. In Table 1, the coordinates, apparent visual magnitudes, and spectral types of the variable and the comparison stars are given. Although the variable and comparison stars are very close in the sky, differential atmospheric extinction corrections were applied. The atmospheric extinction coefficients were obtained from observations of the comparison stars on each night. Heliocentric corrections were also applied to the times of the observations. The mean averages of the standard deviations are 0$^m$.025 for observations acquired in the $V$ band. To compute the standard deviations of observations, we used the standard deviations of the reduced differential magnitudes in the sense comparisons minus the check stars for each night. All KR Cyg data was phased using the minimum time and period taken from \citet{sipahii}. Then, the $V$-band light curve shown in Figure 1 was derived.

\begin{table*}
\begin{center}
\caption{Basic parameters of the observed stars.}\label{T1}
\begin{tabular}{@{}lcccclccrc@{}}
\hline
Star   &  $\alpha_{2025}$ & $\delta_{2025}$ & $V$    & Spectral Type  \\
       &  (hh:mm:ss.ss)   & (dd:mm:ss.ss)   & (mag)  &                \\
\hline
 KR Cyg    & 20:10:05.83 & +30:37:29.60 & 9.34 & A0  \\
 HD 191398 & 20:09:39.73 & +30:24:42.96 & 9.00 & A0V \\
 HD 333664 & 20:10:18.54 & +30:18:07.56 & 9.63 & A0  \\
\hline
\end{tabular}
\end{center}
\end{table*}

KR Cyg eclipsing binary has been observed at different intervals over 26 years at the Ege University Observatory. Discussions on the light variations of the KR Cyg system obtained in different years have been provided by \citet{sipahi}, \citet{sipahit}, \citet{sipahii}, and \citet{sipahiii}. Observations obtained in the $V$ band in 2024 demonstrated that the system's light curves exhibit similar variations. The depths of the minima do not change. Both minima are sufficiently symmetrical. The secondary minimum lies just at the phase 0.5. Since the radial velocity of the system, the solution of the light curve, and the absolute parameters are provided in \citet{sipahiii}, the parameters of that study have been adopted in this work.

\begin{figure*}
{\includegraphics[width=15cm]{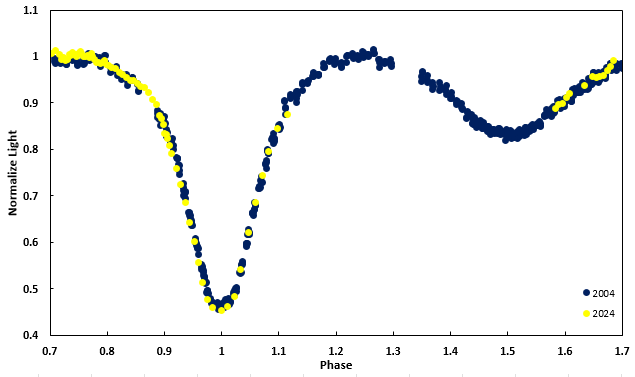}}
\caption{The $V$-band normalized light curves of the KR Cyg from 2004 (blue-coloured points) and 2024 (yellow-coloured points) are presented (brightness values were converted to intensity to compare the light curve variations over a 20-year time interval).}
\label{F1}
\end{figure*}

\section{Orbital Period Variation}
To investigate the orbital period variation of KR Cyg, all accessible minimum times were gathered from the literature. In addition, the system was observed for three nights in 2024. From these observations, a primary minimum time was determined. Additionally, the primary and secondary minimum times of the system have been determined from the Transiting Exoplanet Survey Satellite ({\it TESS}) data \citep{Ricker_2015}. All other minimum times were obtained from the database of (O-C) Gateway \citep{pas}. All minimum timings of KR Cyg eclipsing binary are provided in the appendix. In total, 179 visual, 29 photographic, and 219 photoelectric minimum times of KR Cyg were used in the period analysis. The (O-C) residuals of KR Cyg were calculated using the following linear ephemeris and plotted in Figure 2. 
\begin{equation}
JD~({\rm Hel.})~=~2455036.5310(14)~+~0^{d}.84515279(6)~\times~E.
\end{equation}

\begin{figure*}
{\includegraphics[width=15cm]{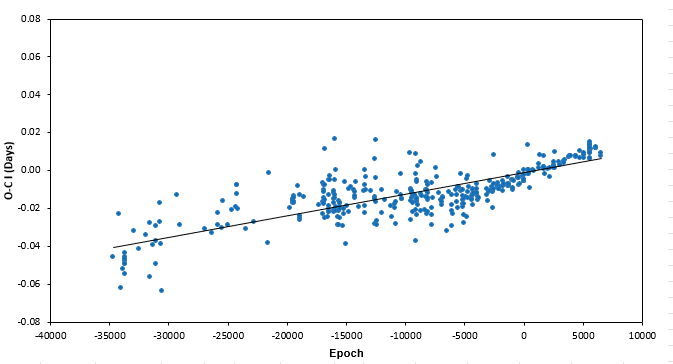}}
\caption{(O-C) diagram for KR Cyg.}
\label{F1}
\end{figure*}

\begin{figure*}
{\includegraphics[width=15cm]{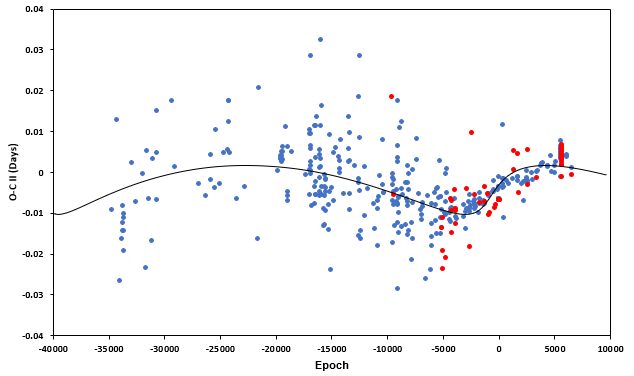}}
\caption{Residuals (O-C II) for KR Cyg. The filled blue and red circles represent the primary and secondary minima, respectively.}
\label{F1}
\end{figure*}

Subsequently, the (O-C)I values and (O-C)II residuals were plotted against the epoch number ($E$) in Figure 2 and Figure 3, respectively. The residuals (O-C) II suggest that the orbital period of KR Cyg exhibits a long-term sinusoidal variation. It is hypothesized that the most plausible explanation for the cyclic variation in the (O-C)II residuals is the light-travel-time effect (LTTE) caused by an undetected third body in the system. To model this effect, the LTTE equation, presented as expression (2) by \citet{Irwin1959}, was applied to the (O-C)II residuals of the system's eclipse timings.
\begin{equation}
\Delta t~=~\frac{a_{12}\sin i'}{c} \Biggl\{\frac{1-e'^{2}}{1+e'\cos\nu'}\sin(\nu'+\omega')+e'\cos\omega'\Biggl\}
\end{equation}
Here, $\Delta t$ represents the time shift caused by the LTTE, $c$ is the speed of light, and $a_{12}$, $i'$, $e'$and $\omega'$ denote the semi-major axis, inclination, eccentricity, and longitude of periastron of the absolute orbit of the eclipsing binary’s center of mass around the triple system’s center of mass. Additionally, $\nu'$ refers to the true anomaly of the eclipsing binary's center of mass along this orbit.

A weighted least-squares analysis was performed to determine two parameters ($T_{0}$ and $P$) for the linear ephemeris of KR Cyg, along with five parameters ($a_{12} \sin i'$, $e'$, $\omega'$, $T'$ and $P'$) for the LTTE. The results of this analysis are summarized in Table 2. The observed data points, along with the theoretical curve providing the best fit, are plotted against the epoch number in Figure 3. Assuming that the orbit of the presumed third body is circular, we obtained the mass function as $f(m) = 0.00157M_{\odot}$ for the third body, using the Equation:
\begin{equation}
f(m)~=~\frac{4\pi^{2}}{GT^{2}}\times(a_{12}\sin i')^{3}=\frac{(M_{3}~\sin i')^{3}}{(M_{1}+ M_{2}+ M_{3})^{2}}
\end{equation}
where $M_1$, $M_2$, and $M_3$ are the masses of the binary's components and the third body, respectively. 

The mass of the third component can be determined using Equation (3), which varies based on the orbital inclination. For instance, when $i'=90^\circ$, the minimum mass $M_{{\rm 3,min}}$ is calculated to be 0.32 $M_{\odot}$. In this calculation, the total mass of the eclipsing binary system is assumed to be 4.14 $M_{\odot}$. The masses of the third body for various inclination angles $i'$  have been computed and are presented in Table 3. The data in Table 3 also indicate that KR Cyg follows an eccentric orbit around the center of mass of the triple system, with a period of approximately $\sim$71 years.

\begin{table*}
\begin{center}
\caption{The parameters derived from (O-C) analysis of KR Cyg.}\label{T1}
\begin{tabular}{@{}lccrrlccrr@{}}
\hline
Parameter & Value \\
\hline
$T_0$ (HJD) & 2455036.5310±0.0014 \\
 P (day) &  0.84515279±0.00000006 \\
 K (day) &   0.012±0.006 \\
 $P^\prime$ (year) &  70.9±1.9 \\
 $e^\prime$ & 0.93±0.12 \\
 $\omega^\prime$ (deg) &  26.70±0.02 \\
 ${a_{12} \sin i^\prime}$ (AU) & 1.99±0.22 \\
 $f(m)$ ($M_{\odot}$) &  0.00157±0.00010 \\
 $M_3$ ($i^\prime$=15\textdegree) & 1.41$M_{\odot}$ \\
 $M_3$ ($i^\prime$=30\textdegree) & 0.66$M_{\odot}$ \\
 $M_3$ ($i^\prime$=45\textdegree) & 0.45$M_{\odot}$ \\
 $M_3$ ($i^\prime$=60\textdegree) & 0.36$M_{\odot}$ \\
 $M_3$ ($i^\prime$=75\textdegree) & 0.33$M_{\odot}$ \\
 $M_3$ ($i^\prime$=90\textdegree) & 0.32$M_{\odot}$ \\
\hline
\end{tabular}
\end{center}
\end{table*}

\section{Conclusion}
In this study, the orbital period variation of the KR Cyg system was reanalysed using data based on its 2024 photometry. The system has been observed at Ege University observatory during certain years and monitored for approximately 26 years. During this time, photometric and spectroscopic studies of the system were obtained, and the absolute parameters of the components were determined by \citet{sipahiii}. The orbital period variation caused by a third body was proposed by \citet{sipahii}. All minimum time observations conducted after the period variation study proposed by \citet{sipahii} have supported this suggestion. The (O-C) variation of the KR Cyg appears scattered. Both the primary and secondary minima exhibit the same variation in the (O-C) diagram. With our 2024 observations, the system's period variation has been updated under the assumption that the orbital period change is caused by a third body. \citet{tvar} also examined the period variation of the KR Cyg system and determined the orbital period of the third body to be approximately 80 years. The minimum timings to be obtained in the next approximately five years are 
crucial for refining the orbital parameters of the third body of the system. 

The latest data obtained for the eclipsing binary KR Cyg indicate that there is no significant change in the light curves. Detailed analyses of the characteristics of the light curves of the system have been provided by \citet{sipahii}. According to these authors, comparison of the radii of the components with their corresponding Roche lobes suggests that the KR Cyg system is near-contact but not in contact. By combining the results of light curve and radial velocity analyses, they determined the absolute physical parameters of the system. Both components are located on the main sequence of the Hertzsprung-Russell diagram. The systematic behaviour of the residuals between the observed and fitted light curves is attributed to the mutual heating effect of the components, particularly in the case of the cooler star in this near-contact system. Analyses indicate that the effective albedo and effective temperature of the irradiated star are significantly altered. The empirically derived albedo value is smaller than expected for a convective star. As the albedo decreases, the effective temperature increases.

Near-contact binary systems are of significant interest in astrophysics due to their role in understanding mass transfer, angular momentum evolution, and stellar mergers. Just like KR Cyg eclipsing binary system, V994 Herculis, RZ Dra, and CN And illustrate the diversity among NCB systems with a third component. In V994 Her eclipsing binary system, the presence of a tertiary component has been confirmed through spectroscopic analysis, revealing a hierarchical triple system \citep{zas}. The third body of the system influences the orbital evolution of the binary and contributes to its mass-transfer dynamics. Similarly, RZ Dra, a well-studied eclipsing binary, shows evidence of long-term orbital period variations, attributed to a distant third companion \citep{erdem}. Meanwhile, CN And stands out due to its compact tertiary component, which significantly perturbs the inner binary's orbit, creating detectable eclipsing light curve variations \citep{cai}. These systems provide critical insights into the dynamical stability of multi-star systems and the mechanisms driving their evolution. Observations of such binaries with a third component challenge existing models of stellar evolution, particularly regarding angular momentum redistribution and the eventual fate of such systems, which may evolve into contact binaries or merge into single stars.

From these investigations, we could draw out the following conclusions: 
\begin{itemize}
\item The orbital period variation of the system is caused by a third body. According to the results of the period variation analysis, the orbital period of the third body has been updated in $\sim$71 years.
\item No significant changes have been observed in the light curves of the system over approximately 26 years of observations.
\item High-resolution spectroscopic observations are required to test the existence of the third body in the system.
\item Observations of minimum timings to be obtained in the coming years are crucial for determining the orbital parameters of the third body more precisely. Observations should continue.
\end{itemize}

\section*{Acknowledgements}
This study was supported by TÜBİTAK (Scientific and Technological Research Council of Turkey) under Grant No. TÜBİTAK-2209A-1919B012336018. The authors wish to thank all the staff of the Ege University Observatory for the allocation of telescope time.




\section*{Supplementary}

\begin{center}
Times of minima of KR Cyg.
\end{center}

\renewcommand{\arraystretch}{0.4}
\setlength{\tabcolsep}{6pt}

\begin{longtable}{|r|r|r|r|l|l|}
\hline
\textbf{HJD (+24 00000)} & \textbf{Year} & \textbf{Cycle} & \textbf{(O-C)} & \textbf{Method} & \textbf{References} \\
\hline
\endfirsthead

\hline
\textbf{HJD (+24 00000)} & \textbf{Year} & \textbf{Cycle} & \textbf{(O-C)} & \textbf{Method} & \textbf{References} \\
\hline
\endhead

\hline
\endfoot

25700.4230	&	1929.3	&	-34711.0	&	-0.04594	&	pg	&	O-C Gateway	\\
26120.4860	&	1930.5	&	-34214.0	&	-0.02334	&	pg	&	O-C Gateway	\\
26298.7740	&	1931.0	&	-34003.0	&	-0.06234	&	pg	&	O-C Gateway	\\
26447.5310	&	1931.4	&	-33827.0	&	-0.05204	&	pg	&	O-C Gateway	\\
26586.1400	&	1931.8	&	-33663.0	&	-0.04792	&	vis	&	O-C Gateway	\\
26590.3700	&	1931.8	&	-33658.0	&	-0.04368	&	vis	&	O-C Gateway	\\
26591.2090	&	1931.8	&	-33657.0	&	-0.04983	&	vis	&	O-C Gateway	\\
26596.2800	&	1931.8	&	-33651.0	&	-0.04974	&	vis	&	O-C Gateway	\\
26601.3550	&	1931.8	&	-33645.0	&	-0.04565	&	vis	&	O-C Gateway	\\
26602.1910	&	1931.8	&	-33644.0	&	-0.05481	&	vis	&	O-C Gateway	\\
26607.2700	&	1931.8	&	-33638.0	&	-0.04672	&	vis	&	O-C Gateway	\\
27245.3740	&	1933.6	&	-32883.0	&	-0.03225	&	vis	&	O-C Gateway	\\
27553.0000	&	1934.4	&	-32519.0	&	-0.04147	&	V	&	O-C Gateway	\\
28074.4660	&	1935.8	&	-31902.0	&	-0.03407	&	pg	&	O-C Gateway	\\
28333.9050	&	1936.5	&	-31595.0	&	-0.05664	&	V	&	O-C Gateway	\\
28380.4170	&	1936.7	&	-31540.0	&	-0.02798	&	pg	&	O-C Gateway	\\
28545.2100	&	1937.1	&	-31345.0	&	-0.03956	&	pg	&	O-C Gateway	\\
28773.3910	&	1937.7	&	-31075.0	&	-0.04952	&	pg	&	O-C Gateway	\\
28806.3640	&	1937.8	&	-31036.0	&	-0.03744	&	pg	&	O-C Gateway	\\
28811.4430	&	1937.8	&	-31030.0	&	-0.02935	&	pg	&	O-C Gateway	\\
29106.4027	&	1938.7	&	-30681.0	&	-0.02759	&	pg	&	O-C Gateway	\\
29106.4130	&	1938.7	&	-30681.0	&	-0.01729	&	pg	&	O-C Gateway	\\
29143.5780	&	1938.8	&	-30637.0	&	-0.03897	&	pg	&	O-C Gateway	\\
29230.6040	&	1939.0	&	-30534.0	&	-0.06359	&	pg	&	O-C Gateway	\\
30259.2040	&	1941.8	&	-29317.0	&	-0.01321	&	pg	&	O-C Gateway	\\
30531.3270	&	1942.6	&	-28995.0	&	-0.02906	&	pg	&	O-C Gateway	\\
32296.8470	&	1947.4	&	-26906.0	&	-0.03096	&	pg	&	O-C Gateway	\\
32821.6840	&	1948.8	&	-26285.0	&	-0.03317	&	pg	&	O-C Gateway	\\
33220.6060	&	1949.9	&	-25813.0	&	-0.02277	&	pg	&	O-C Gateway	\\
33226.5160	&	1949.9	&	-25806.0	&	-0.02883	&	pg	&	O-C Gateway	\\
33536.6850	&	1950.8	&	-25439.0	&	-0.0305	&	pg	&	O-C Gateway	\\
33567.9700	&	1950.9	&	-25402.0	&	-0.01612	&	pg	&	O-C Gateway	\\
33923.7660	&	1951.8	&	-24981.0	&	-0.02898	&	pg	&	O-C Gateway	\\
34213.6610	&	1952.6	&	-24638.0	&	-0.02102	&	pg	&	O-C Gateway	\\
34512.8460	&	1953.5	&	-24284.0	&	-0.01972	&	pe	&	O-C Gateway	\\
34596.5280	&	1953.7	&	-24185.0	&	-0.00774	&	vis	&	O-C Gateway	\\
34602.4390	&	1953.7	&	-24178.0	&	-0.0128	&	vis	&	O-C Gateway	\\
34607.5150	&	1953.7	&	-24172.0	&	-0.00771	&	vis	&	O-C Gateway	\\
34650.6050	&	1953.8	&	-24121.0	&	-0.02044	&	pg	&	O-C Gateway	\\
35221.9170	&	1955.4	&	-23445.0	&	-0.03099	&	pg	&	O-C Gateway	\\
35774.6500	&	1956.9	&	-22791.0	&	-0.02721	&	pg	&	O-C Gateway	\\
36818.4010	&	1959.8	&	-21556.0	&	-0.03855	&	pg	&	O-C Gateway	\\
36868.3020	&	1959.9	&	-21497.0	&	-0.00151	&	pg	&	O-C Gateway	\\
38336.3120	&	1963.9	&	-19760.0	&	-0.02001	&	vis	&	O-C Gateway	\\
38558.5900	&	1964.5	&	-19497.0	&	-0.01691	&	pe	&	O-C Gateway	\\
38559.4352	&	1964.5	&	-19496.0	&	-0.01686	&	pe	&	O-C Gateway	\\
38559.4360	&	1964.5	&	-19496.0	&	-0.01606	&	pe	&	O-C Gateway	\\
38580.5649	&	1964.6	&	-19471.0	&	-0.01595	&	pe	&	O-C Gateway	\\
38580.5657	&	1964.6	&	-19471.0	&	-0.01515	&	pe	&	O-C Gateway	\\
38597.4671	&	1964.6	&	-19451.0	&	-0.01678	&	pe	&	O-C Gateway	\\
38608.4535	&	1964.7	&	-19438.0	&	-0.01736	&	pe	&	O-C Gateway	\\
38614.3693	&	1964.7	&	-19431.0	&	-0.01762	&	pe	&	O-C Gateway	\\
38614.3702	&	1964.7	&	-19431.0	&	-0.01672	&	pe	&	O-C Gateway	\\
38652.4050	&	1964.8	&	-19386.0	&	-0.01374	&	vis	&	O-C Gateway	\\
38675.2210	&	1964.9	&	-19359.0	&	-0.01684	&	pe	&	O-C Gateway	\\
38675.2229	&	1964.9	&	-19359.0	&	-0.01494	&	pe	&	O-C Gateway	\\
38941.4520	&	1965.6	&	-19044.0	&	-0.00863	&	vis	&	O-C Gateway	\\
39040.3170	&	1965.9	&	-18927.0	&	-0.02637	&	vis	&	O-C Gateway	\\
39040.3300	&	1965.9	&	-18927.0	&	-0.01337	&	vis	&	O-C Gateway	\\
39051.3050	&	1965.9	&	-18914.0	&	-0.02535	&	vis	&	O-C Gateway	\\
39051.3060	&	1965.9	&	-18914.0	&	-0.02435	&	vis	&	O-C Gateway	\\
39389.3770	&	1966.8	&	-18514.0	&	-0.01403	&	vis	&	O-C Gateway	\\
40420.4578	&	1969.6	&	-17294.0	&	-0.0183	&	vis	&	O-C Gateway	\\
40725.5590	&	1970.5	&	-16933.0	&	-0.01686	&	vis	&	O-C Gateway	\\
40731.4690	&	1970.5	&	-16926.0	&	-0.02293	&	vis	&	O-C Gateway	\\
40753.4590	&	1970.5	&	-16900.0	&	-0.00687	&	vis	&	O-C Gateway	\\
40759.3740	&	1970.6	&	-16893.0	&	-0.00793	&	vis	&	O-C Gateway	\\
40780.5000	&	1970.6	&	-16868.0	&	-0.01072	&	vis	&	O-C Gateway	\\
40785.5700	&	1970.6	&	-16862.0	&	-0.01163	&	vis	&	O-C Gateway	\\
40786.4100	&	1970.6	&	-16861.0	&	-0.01679	&	vis	&	O-C Gateway	\\
40791.4800	&	1970.7	&	-16855.0	&	-0.0177	&	vis	&	O-C Gateway	\\
40796.5540	&	1970.7	&	-16849.0	&	-0.01461	&	vis	&	O-C Gateway	\\
40830.3670	&	1970.8	&	-16809.0	&	-0.00767	&	vis	&	O-C Gateway	\\
40830.3860	&	1970.8	&	-16809.0	&	0.01133	&	vis	&	O-C Gateway	\\
40841.3460	&	1970.8	&	-16796.0	&	-0.01565	&	pe	&	O-C Gateway	\\
40890.3550	&	1970.9	&	-16738.0	&	-0.02545	&	vis	&	O-C Gateway	\\
41080.5150	&	1971.4	&	-16513.0	&	-0.02458	&	vis	&	O-C Gateway	\\
41135.4540	&	1971.6	&	-16448.0	&	-0.02044	&	vis	&	O-C Gateway	\\
41146.4560	&	1971.6	&	-16435.0	&	-0.00541	&	vis	&	O-C Gateway	\\
41162.4970	&	1971.7	&	-16416.0	&	-0.02229	&	vis	&	O-C Gateway	\\
41168.4130	&	1971.7	&	-16409.0	&	-0.02235	&	vis	&	O-C Gateway	\\
41173.4930	&	1971.7	&	-16403.0	&	-0.01326	&	vis	&	O-C Gateway	\\
41174.3323	&	1971.7	&	-16402.0	&	-0.01912	&	pe	&	O-C Gateway	\\
41201.3910	&	1971.8	&	-16370.0	&	-0.00527	&	vis	&	O-C Gateway	\\
41234.3540	&	1971.9	&	-16331.0	&	-0.00319	&	vis	&	O-C Gateway	\\
41490.4170	&	1972.6	&	-16028.0	&	-0.02115	&	vis	&	O-C Gateway	\\
41490.4230	&	1972.6	&	-16028.0	&	-0.01515	&	vis	&	O-C Gateway	\\
41506.4740	&	1972.6	&	-16009.0	&	-0.02203	&	vis	&	O-C Gateway	\\
41511.5550	&	1972.6	&	-16003.0	&	-0.01194	&	vis	&	O-C Gateway	\\
41522.5430	&	1972.7	&	-15990.0	&	-0.01092	&	vis	&	O-C Gateway	\\
41528.4524	&	1972.7	&	-15983.0	&	-0.01758	&	vis	&	O-C Gateway	\\
41556.3550	&	1972.7	&	-15950.0	&	-0.00499	&	vis	&	O-C Gateway	\\
41583.3900	&	1972.8	&	-15918.0	&	-0.01484	&	vis	&	O-C Gateway	\\
41589.3080	&	1972.8	&	-15911.0	&	-0.0129	&	vis	&	O-C Gateway	\\
41594.4080	&	1972.8	&	-15905.0	&	0.01619	&	vis	&	O-C Gateway	\\
41599.4410	&	1972.9	&	-15899.0	&	-0.02172	&	vis	&	O-C Gateway	\\
41622.2820	&	1972.9	&	-15872.0	&	0.00018	&	vis	&	O-C Gateway	\\
41627.3310	&	1972.9	&	-15866.0	&	-0.02173	&	B	&	O-C Gateway	\\
41627.3330	&	1972.9	&	-15866.0	&	-0.01973	&	vis	&	O-C Gateway	\\
41812.4120	&	1973.4	&	-15647.0	&	-0.02895	&	vis	&	O-C Gateway	\\
41823.4030	&	1973.5	&	-15634.0	&	-0.02492	&	vis	&	O-C Gateway	\\
41850.4440	&	1973.5	&	-15602.0	&	-0.02878	&	vis	&	O-C Gateway	\\
41850.4510	&	1973.5	&	-15602.0	&	-0.02178	&	vis	&	O-C Gateway	\\
41871.5850	&	1973.6	&	-15577.0	&	-0.01657	&	vis	&	O-C Gateway	\\
41904.5430	&	1973.7	&	-15538.0	&	-0.01949	&	vis	&	O-C Gateway	\\
41931.5820	&	1973.8	&	-15506.0	&	-0.02534	&	vis	&	O-C Gateway	\\
41932.4330	&	1973.8	&	-15505.0	&	-0.01949	&	vis	&	O-C Gateway	\\
41938.3480	&	1973.8	&	-15498.0	&	-0.02055	&	vis	&	O-C Gateway	\\
41938.3510	&	1973.8	&	-15498.0	&	-0.01755	&	vis	&	O-C Gateway	\\
41943.4220	&	1973.8	&	-15492.0	&	-0.01746	&	vis	&	O-C Gateway	\\
42144.5560	&	1974.4	&	-15254.0	&	-0.02957	&	vis	&	O-C Gateway	\\
42193.5810	&	1974.5	&	-15196.0	&	-0.02337	&	vis	&	O-C Gateway	\\
42221.4741	&	1974.6	&	-15163.0	&	-0.02027	&	vis	&	O-C Gateway	\\
42254.4490	&	1974.7	&	-15124.0	&	-0.00629	&	vis	&	O-C Gateway	\\
42337.2410	&	1974.9	&	-15026.0	&	-0.03916	&	vis	&	O-C Gateway	\\
42419.2410	&	1975.1	&	-14929.0	&	-0.01887	&	vis	&	O-C Gateway	\\
42570.5200	&	1975.5	&	-14750.0	&	-0.02203	&	vis	&	O-C Gateway	\\
42576.4480	&	1975.5	&	-14743.0	&	-0.01009	&	vis	&	O-C Gateway	\\
42708.2930	&	1975.9	&	-14587.0	&	-0.00875	&	vis	&	O-C Gateway	\\
42958.4560	&	1976.6	&	-14291.0	&	-0.01066	&	vis	&	O-C Gateway	\\
42980.4260	&	1976.6	&	-14265.0	&	-0.0146	&	vis	&	O-C Gateway	\\
42996.4850	&	1976.7	&	-14246.0	&	-0.01348	&	vis	&	O-C Gateway	\\
43013.3900	&	1976.7	&	-14226.0	&	-0.01152	&	vis	&	O-C Gateway	\\
43046.3560	&	1976.8	&	-14187.0	&	-0.00643	&	vis	&	O-C Gateway	\\
43280.4500	&	1977.5	&	-13910.0	&	-0.01945	&	vis	&	O-C Gateway	\\
43673.4540	&	1978.5	&	-13445.0	&	-0.01099	&	vis	&	O-C Gateway	\\
43689.5150	&	1978.6	&	-13426.0	&	-0.00788	&	vis	&	O-C Gateway	\\
43706.4260	&	1978.6	&	-13406.0	&	0.00009	&	vis	&	O-C Gateway	\\
43739.3680	&	1978.7	&	-13367.0	&	-0.01883	&	vis	&	O-C Gateway	\\
43744.4350	&	1978.7	&	-13361.0	&	-0.02274	&	vis	&	O-C Gateway	\\
43766.4210	&	1978.8	&	-13335.0	&	-0.01068	&	vis	&	O-C Gateway	\\
43777.4150	&	1978.8	&	-13322.0	&	-0.00365	&	vis	&	O-C Gateway	\\
44165.3320	&	1979.9	&	-12863.0	&	-0.01128	&	vis	&	O-C Gateway	\\
44426.4810	&	1980.6	&	-12554.0	&	-0.01416	&	vis	&	O-C Gateway	\\
44437.4880	&	1980.6	&	-12541.0	&	0.00587	&	vis	&	O-C Gateway	\\
44443.3700	&	1980.6	&	-12534.0	&	-0.02819	&	vis	&	O-C Gateway	\\
44454.3700	&	1980.7	&	-12521.0	&	-0.01516	&	vis	&	O-C Gateway	\\
44498.3490	&	1980.8	&	-12469.0	&	0.01595	&	vis	&	O-C Gateway	\\
44503.3870	&	1980.8	&	-12463.0	&	-0.01696	&	vis	&	O-C Gateway	\\
44514.3870	&	1980.8	&	-12450.0	&	-0.00394	&	vis	&	O-C Gateway	\\
44569.2970	&	1981.0	&	-12385.0	&	-0.0288	&	vis	&	O-C Gateway	\\
44569.2990	&	1981.0	&	-12385.0	&	-0.0268	&	vis	&	O-C Gateway	\\
44913.2800	&	1981.9	&	-11978.0	&	-0.02254	&	vis	&	O-C Gateway	\\
44913.2920	&	1981.9	&	-11978.0	&	-0.01054	&	vis	&	O-C Gateway	\\
45130.4907	&	1982.5	&	-11721.0	&	-0.01582	&	vis	&	O-C Gateway	\\
45600.3920	&	1983.8	&	-11165.0	&	-0.01887	&	vis	&	O-C Gateway	\\
45644.3340	&	1983.9	&	-11113.0	&	-0.02476	&	vis	&	O-C Gateway	\\
45878.4480	&	1984.6	&	-10836.0	&	-0.01778	&	vis	&	O-C Gateway	\\
45889.4330	&	1984.6	&	-10823.0	&	-0.01975	&	vis	&	O-C Gateway	\\
45911.3980	&	1984.7	&	-10797.0	&	-0.0287	&	vis	&	O-C Gateway	\\
45933.3840	&	1984.7	&	-10771.0	&	-0.01664	&	vis	&	O-C Gateway	\\
46004.3820	&	1984.9	&	-10687.0	&	-0.01138	&	vis	&	O-C Gateway	\\
46271.4480	&	1985.7	&	-10371.0	&	-0.01332	&	vis	&	O-C Gateway	\\
46271.4590	&	1985.7	&	-10371.0	&	-0.00232	&	vis	&	O-C Gateway	\\
46326.3810	&	1985.8	&	-10306.0	&	-0.01518	&	vis	&	O-C Gateway	\\
46331.4520	&	1985.8	&	-10300.0	&	-0.01509	&	vis	&	O-C Gateway	\\
46917.5890	&	1987.4	&	-9606.5	&	0.00921	&	pe	&	O-C Gateway	\\
46992.3660	&	1987.6	&	-9518.0	&	-0.00972	&	vis	&	O-C Gateway	\\
47003.3520	&	1987.7	&	-9505.0	&	-0.01069	&	vis	&	O-C Gateway	\\
47008.4070	&	1987.7	&	-9499.0	&	-0.0266	&	vis	&	O-C Gateway	\\
47023.6300	&	1987.7	&	-9481.0	&	-0.01633	&	vis	&	O-C Gateway	\\
47039.6920	&	1987.8	&	-9462.0	&	-0.01221	&	vis	&	O-C Gateway	\\
47055.3250	&	1987.8	&	-9443.5	&	-0.01452	&	vis	&	O-C Gateway	\\
47083.6410	&	1987.9	&	-9410.0	&	-0.0111	&	vis	&	O-C Gateway	\\
47111.5250	&	1988.0	&	-9377.0	&	-0.01711	&	vis	&	O-C Gateway	\\
47121.6700	&	1988.0	&	-9365.0	&	-0.01393	&	vis	&	O-C Gateway	\\
47151.2420	&	1988.1	&	-9330.0	&	-0.02224	&	vis	&	O-C Gateway	\\
47300.8430	&	1988.5	&	-9153.0	&	-0.01309	&	vis	&	O-C Gateway	\\
47362.5360	&	1988.6	&	-9080.0	&	-0.01616	&	pe	&	O-C Gateway	\\
47368.4310	&	1988.7	&	-9073.0	&	-0.03723	&	vis	&	O-C Gateway	\\
47374.3630	&	1988.7	&	-9066.0	&	-0.02129	&	vis	&	O-C Gateway	\\
47374.3820	&	1988.7	&	-9066.0	&	-0.00229	&	vis	&	O-C Gateway	\\
47385.3800	&	1988.7	&	-9053.0	&	0.00874	&	vis	&	O-C Gateway	\\
47407.3400	&	1988.8	&	-9027.0	&	-0.0052	&	vis	&	O-C Gateway	\\
47410.7090	&	1988.8	&	-9023.0	&	-0.01681	&	vis	&	O-C Gateway	\\
47412.3997	&	1988.8	&	-9021.0	&	-0.01641	&	vis	&	O-C Gateway	\\
47423.3970	&	1988.8	&	-9008.0	&	-0.00609	&	vis	&	O-C Gateway	\\
47448.7340	&	1988.9	&	-8978.0	&	-0.02364	&	vis	&	O-C Gateway	\\
47460.5750	&	1988.9	&	-8964.0	&	-0.01476	&	vis	&	O-C Gateway	\\
47462.2650	&	1988.9	&	-8962.0	&	-0.01506	&	vis	&	O-C Gateway	\\
47498.6030	&	1989.0	&	-8919.0	&	-0.01859	&	vis	&	O-C Gateway	\\
47522.2880	&	1989.1	&	-8891.0	&	0.00216	&	vis	&	O-C Gateway	\\
47712.4325	&	1989.6	&	-8666.0	&	-0.01244	&	U	&	O-C Gateway	\\
47721.7200	&	1989.6	&	-8655.0	&	-0.02164	&	vis	&	O-C Gateway	\\
47734.4230	&	1989.7	&	-8640.0	&	0.00409	&	vis	&	O-C Gateway	\\
47805.3980	&	1989.9	&	-8556.0	&	-0.01365	&	vis	&	O-C Gateway	\\
48012.4530	&	1990.4	&	-8311.0	&	-0.02082	&	vis	&	O-C Gateway	\\
48043.7340	&	1990.5	&	-8274.0	&	-0.01043	&	vis	&	O-C Gateway	\\
48065.7040	&	1990.6	&	-8248.0	&	-0.01438	&	vis	&	O-C Gateway	\\
48127.4040	&	1990.7	&	-8175.0	&	-0.01045	&	vis	&	O-C Gateway	\\
48143.4660	&	1990.8	&	-8156.0	&	-0.00633	&	vis	&	O-C Gateway	\\
48158.6620	&	1990.8	&	-8138.0	&	-0.02307	&	vis	&	O-C Gateway	\\
48180.6430	&	1990.9	&	-8112.0	&	-0.01601	&	vis	&	O-C Gateway	\\
48191.6310	&	1990.9	&	-8099.0	&	-0.01498	&	vis	&	O-C Gateway	\\
48202.6110	&	1990.9	&	-8086.0	&	-0.02195	&	vis	&	O-C Gateway	\\
48208.5410	&	1991.0	&	-8079.0	&	-0.00802	&	vis	&	O-C Gateway	\\
48235.5820	&	1991.0	&	-8047.0	&	-0.01187	&	vis	&	O-C Gateway	\\
48469.6810	&	1991.7	&	-7770.0	&	-0.01989	&	vis	&	O-C Gateway	\\
48491.6460	&	1991.7	&	-7744.0	&	-0.02884	&	vis	&	O-C Gateway	\\
48507.7090	&	1991.8	&	-7725.0	&	-0.02372	&	vis	&	O-C Gateway	\\
48535.6160	&	1991.9	&	-7692.0	&	-0.00672	&	vis	&	O-C Gateway	\\
48546.5950	&	1991.9	&	-7679.0	&	-0.0147	&	vis	&	O-C Gateway	\\
48820.4400	&	1992.6	&	-7355.0	&	0.00115	&	vis	&	O-C Gateway	\\
48853.3960	&	1992.7	&	-7316.0	&	-0.00376	&	vis	&	O-C Gateway	\\
48983.5410	&	1993.1	&	-7162.0	&	-0.01212	&	vis	&	O-C Gateway	\\
49164.4000	&	1993.6	&	-6948.0	&	-0.01559	&	vis	&	O-C Gateway	\\
49195.6680	&	1993.7	&	-6911.0	&	-0.0182	&	vis	&	O-C Gateway	\\
49238.7720	&	1993.8	&	-6860.0	&	-0.01694	&	vis	&	O-C Gateway	\\
49266.6650	&	1993.9	&	-6827.0	&	-0.01394	&	vis	&	O-C Gateway	\\
49271.7350	&	1993.9	&	-6821.0	&	-0.01485	&	vis	&	O-C Gateway	\\
49562.4500	&	1994.7	&	-6477.0	&	-0.03204	&	vis	&	O-C Gateway	\\
49687.5530	&	1995.0	&	-6329.0	&	-0.01149	&	vis	&	O-C Gateway	\\
49926.7130	&	1995.7	&	-6046.0	&	-0.02942	&	vis	&	O-C Gateway	\\
49948.6950	&	1995.7	&	-6020.0	&	-0.02137	&	CCD	&	O-C Gateway	\\
49953.7740	&	1995.7	&	-6014.0	&	-0.01328	&	vis	&	O-C Gateway	\\
49958.8400	&	1995.8	&	-6008.0	&	-0.01819	&	pe	&	O-C Gateway	\\
49965.5960	&	1995.8	&	-6000.0	&	-0.0234	&	pe	&	O-C Gateway	\\
50226.7610	&	1996.5	&	-5691.0	&	-0.01028	&	pe	&	O-C Gateway	\\
50266.4764	&	1996.6	&	-5644.0	&	-0.01701	&	CCD	&	O-C Gateway	\\
50308.7380	&	1996.7	&	-5594.0	&	-0.01299	&	vis	&	O-C Gateway	\\
50325.6340	&	1996.8	&	-5574.0	&	-0.02002	&	CCD	&	O-C Gateway	\\
50336.6320	&	1996.8	&	-5561.0	&	-0.009	&	vis	&	O-C Gateway	\\
50402.5540	&	1997.0	&	-5483.0	&	-0.00883	&	vis	&	O-C Gateway	\\
50542.8470	&	1997.3	&	-5317.0	&	-0.01101	&	vis	&	O-C Gateway	\\
50553.8430	&	1997.4	&	-5304.0	&	-0.00198	&	vis	&	O-C Gateway	\\
50668.7750	&	1997.7	&	-5168.0	&	-0.01061	&	vis	&	O-C Gateway	\\
50690.7400	&	1997.8	&	-5142.0	&	-0.01956	&	vis	&	O-C Gateway	\\
50695.8200	&	1997.8	&	-5136.0	&	-0.01047	&	vis	&	O-C Gateway	\\
50696.6600	&	1997.8	&	-5135.0	&	-0.01562	&	vis	&	O-C Gateway	\\
50700.4609	&	1997.8	&	-5130.5	&	-0.0179	&	CCD	&	O-C Gateway	\\
50703.4230	&	1997.8	&	-5127.0	&	-0.01383	&	CCD	&	O-C Gateway	\\
50750.3193	&	1997.9	&	-5071.5	&	-0.02345	&	CCD	&	O-C Gateway	\\
50753.2828	&	1997.9	&	-5068.0	&	-0.01798	&	CCD	&	O-C Gateway	\\
50755.3856	&	1997.9	&	-5065.5	&	-0.02806	&	CCD	&	O-C Gateway	\\
50772.3013	&	1998.0	&	-5045.5	&	-0.0154	&	CCD	&	O-C Gateway	\\
50948.5166	&	1998.5	&	-4837.0	&	-0.01423	&	CCD	&	O-C Gateway	\\
51007.6870	&	1998.6	&	-4767.0	&	-0.00445	&	vis	&	O-C Gateway	\\
51033.4436	&	1998.7	&	-4736.5	&	-0.02497	&	CCD	&	O-C Gateway	\\
51036.4121	&	1998.7	&	-4733.0	&	-0.0145	&	CCD	&	O-C Gateway	\\
51040.6410	&	1998.7	&	-4728.0	&	-0.01136	&	vis	&	O-C Gateway	\\
51083.7520	&	1998.8	&	-4677.0	&	-0.0031	&	vis	&	O-C Gateway	\\
51133.6100	&	1999.0	&	-4618.0	&	-0.00905	&	vis	&	O-C Gateway	\\
51325.4531	&	1999.5	&	-4391.0	&	-0.01539	&	CCD	&	O-C Gateway	\\
51363.4866	&	1999.6	&	-4346.0	&	-0.01371	&	BV	&	\citet{sipahi}	\\
51391.3746	&	1999.7	&	-4313.0	&	-0.01572	&	CCD	&	O-C Gateway	\\
51393.4932	&	1999.7	&	-4310.5	&	-0.01	&	CCD	&	O-C Gateway	\\
51429.4095	&	1999.8	&	-4268.0	&	-0.01264	&	BV	&	\citet{sipahi}	\\
51434.4798	&	1999.8	&	-4262.0	&	-0.01325	&	B	&	O-C Gateway	\\
51443.3488	&	1999.8	&	-4251.5	&	-0.01835	&	CCD	&	O-C Gateway	\\
51454.3437	&	1999.8	&	-4238.5	&	-0.01042	&	BV	&	\citet{sipahi}	\\
51459.4121	&	1999.9	&	-4232.5	&	-0.01293	&	CCD	&	O-C Gateway	\\
51468.2868	&	1999.9	&	-4222.0	&	-0.01232	&	CCD	&	O-C Gateway	\\
51691.4093	&	2000.5	&	-3958.0	&	-0.00987	&	BV	&	\citet{sipahi}	\\
51718.4518	&	2000.6	&	-3926.0	&	-0.01223	&	BV	&	\citet{sipahi}	\\
51721.4147	&	2000.6	&	-3922.5	&	-0.00736	&	BV	&	\citet{sipahi}	\\
51726.4804	&	2000.6	&	-3916.5	&	-0.01257	&	BV	&	\citet{sipahi}	\\
51737.4680	&	2000.6	&	-3903.5	&	-0.01194	&	BV	&	\citet{sipahi}	\\
51797.4700	&	2000.8	&	-3832.5	&	-0.01571	&	CCD	&	O-C Gateway	\\
51816.4875	&	2000.8	&	-3810.0	&	-0.01412	&	CCD	&	O-C Gateway	\\
51850.2932	&	2000.9	&	-3770.0	&	-0.01449	&	CCD	&	O-C Gateway	\\
52411.4782	&	2002.5	&	-3106.0	&	-0.01022	&	-Ir	&	O-C Gateway	\\
52427.5330	&	2002.5	&	-3087.0	&	-0.0133	&	V	&	O-C Gateway	\\
52531.4839	&	2002.8	&	-2964.0	&	-0.01606	&	CCD	&	O-C Gateway	\\
52576.2812	&	2002.9	&	-2911.0	&	-0.0118	&	CCD	&	O-C Gateway	\\
52612.6234	&	2003.0	&	-2868.0	&	-0.01112	&	V	&	O-C Gateway	\\
52613.0512	&	2003.0	&	-2867.5	&	-0.0059	&	V	&	O-C Gateway	\\
52815.4619	&	2003.6	&	-2628.0	&	-0.00903	&	V	&	O-C Gateway	\\
52832.3635	&	2003.6	&	-2608.0	&	-0.01047	&	pe	&	O-C Gateway	\\
52840.3829	&	2003.6	&	-2598.5	&	-0.02001	&	-Ir	&	O-C Gateway	\\
52859.4074	&	2003.7	&	-2576.0	&	-0.01142	&	pe	&	O-C Gateway	\\
52864.4790	&	2003.7	&	-2570.0	&	-0.01073	&	CCD	&	O-C Gateway	\\
52899.1337	&	2003.8	&	-2529.0	&	-0.00725	&	V	&	O-C Gateway	\\
52951.5306	&	2003.9	&	-2467.0	&	-0.00976	&	CCD	&	O-C Gateway	\\
52955.3517	&	2004.0	&	-2462.5	&	0.00816	&	CCD	&	O-C Gateway	\\
53192.4000	&	2004.6	&	-2182.0	&	-0.00859	&	pe	&	O-C Gateway	\\
53195.3598	&	2004.6	&	-2178.5	&	-0.00682	&	UBVR	&	O-C Gateway	\\
53212.6832	&	2004.7	&	-2158.0	&	-0.00903	&	CCD	&	O-C Gateway	\\
53225.3614	&	2004.7	&	-2143.0	&	-0.00811	&	pe	&	O-C Gateway	\\
53267.6169	&	2004.8	&	-2093.0	&	-0.01019	&	CCD	&	O-C Gateway	\\
53269.3084	&	2004.8	&	-2091.0	&	-0.009	&	UBV	&	O-C Gateway	\\
53289.5913	&	2004.9	&	-2067.0	&	-0.00974	&	CCD	&	O-C Gateway	\\
53572.7160	&	2005.6	&	-1732.0	&	-0.01086	&	CCD	&	O-C Gateway	\\
53585.3955	&	2005.7	&	-1717.0	&	-0.00863	&	UBVR	&	O-C Gateway	\\
53593.4249	&	2005.7	&	-1707.5	&	-0.00817	&	UBVR	&	O-C Gateway	\\
53601.4563	&	2005.7	&	-1698.0	&	-0.00571	&	-Ir	&	O-C Gateway	\\
53639.4865	&	2005.8	&	-1653.0	&	-0.00734	&	-Ir	&	O-C Gateway	\\
53891.3416	&	2006.5	&	-1355.0	&	-0.00745	&	uvby	&	O-C Gateway	\\
53907.3985	&	2006.6	&	-1336.0	&	-0.00843	&	pe	&	O-C Gateway	\\
53937.4060	&	2006.6	&	-1300.5	&	-0.00381	&	uvby	&	O-C Gateway	\\
53978.3920	&	2006.8	&	-1252.0	&	-0.00767	&	BV	&	O-C Gateway	\\
53984.3076	&	2006.8	&	-1245.0	&	-0.00813	&	CCD	&	O-C Gateway	\\
53991.4921	&	2006.8	&	-1236.5	&	-0.00742	&	-Ir	&	O-C Gateway	\\
54198.5565	&	2007.4	&	-991.5	&	-0.00519	&	BVRI	&	O-C Gateway	\\
54270.8169	&	2007.6	&	-906.0	&	-0.00526	&	CCD	&	O-C Gateway	\\
54282.2214	&	2007.6	&	-892.5	&	-0.01031	&	V	&	O-C Gateway	\\
54313.4927	&	2007.7	&	-855.5	&	-0.00962	&	-Ir	&	O-C Gateway	\\
54338.4286	&	2007.7	&	-826.0	&	-0.0057	&	V	&	O-C Gateway	\\
54614.7946	&	2008.5	&	-499.0	&	-0.0043	&	CCD	&	O-C Gateway	\\
54653.6721	&	2008.6	&	-453.0	&	-0.00378	&	CCD	&	O-C Gateway	\\
54682.4095	&	2008.7	&	-419.0	&	-0.00156	&	BVR	&	O-C Gateway	\\
54696.3480	&	2008.7	&	-402.5	&	-0.00804	&	B	&	O-C Gateway	\\
54702.6904	&	2008.7	&	-395.0	&	-0.00428	&	CCD	&	O-C Gateway	\\
54762.2706	&	2008.9	&	-324.5	&	-0.00727	&	BVR	&	O-C Gateway	\\
55018.7782	&	2009.6	&	-21.0	&	-0.00321	&	CCD	&	O-C Gateway	\\
55036.5296	&	2009.7	&	0.0	&	0	&	BVR	&	O-C Gateway	\\
55058.4980	&	2009.7	&	26.0	&	-0.00554	&	pe	&	O-C Gateway	\\
55058.9206	&	2009.7	&	26.5	&	-0.00552	&	pe	&	O-C Gateway	\\
55059.3432	&	2009.7	&	27.0	&	-0.0055	&	pe	&	O-C Gateway	\\
55059.7658	&	2009.7	&	27.5	&	-0.00547	&	pe	&	O-C Gateway	\\
55060.1884	&	2009.7	&	28.0	&	-0.00545	&	pe	&	O-C Gateway	\\
55060.6109	&	2009.7	&	28.5	&	-0.00552	&	pe	&	O-C Gateway	\\
55061.0335	&	2009.7	&	29.0	&	-0.0055	&	pe	&	O-C Gateway	\\
55061.4561	&	2009.7	&	29.5	&	-0.00548	&	pe	&	O-C Gateway	\\
55061.8787	&	2009.7	&	30.0	&	-0.00545	&	pe	&	O-C Gateway	\\
55062.3012	&	2009.7	&	30.5	&	-0.00553	&	pe	&	O-C Gateway	\\
55062.7238	&	2009.7	&	31.0	&	-0.0055	&	pe	&	O-C Gateway	\\
55063.1464	&	2009.7	&	31.5	&	-0.00548	&	pe	&	O-C Gateway	\\
55063.5690	&	2009.7	&	32.0	&	-0.00545	&	pe	&	O-C Gateway	\\
55075.4040	&	2009.8	&	46.0	&	-0.00258	&	-Ir	&	O-C Gateway	\\
55096.5329	&	2009.8	&	71.0	&	-0.00247	&	-Ir	&	O-C Gateway	\\
55362.7581	&	2010.5	&	386.0	&	-0.00006	&	V	&	O-C Gateway	\\
55379.6589	&	2010.6	&	406.0	&	-0.00229	&	CCD	&	O-C Gateway	\\
55386.4356	&	2010.6	&	414.0	&	0.0132	&	BVR	&	\citet{sipahii}	\\
55397.4081	&	2010.6	&	427.0	&	-0.00128	&	-Ir	&	O-C Gateway	\\
55419.3832	&	2010.7	&	453.0	&	-0.00012	&	BV	&	\citet{sipahii}	\\
55461.6389	&	2010.8	&	503.0	&	-0.00201	&	V	&	O-C Gateway	\\
55465.0120	&	2010.8	&	507.0	&	-0.00951	&	V	&	O-C Gateway	\\
55691.5220	&	2011.4	&	775.0	&	-0.00017	&	V	&	\citet{sipahii}	\\
55801.3912	&	2011.7	&	905.0	&	-0.00069	&	-Ir	&	O-C Gateway	\\
56147.0599	&	2012.7	&	1314.0	&	0.00097	&	V	&	O-C Gateway	\\
56158.4718	&	2012.7	&	1327.5	&	0.00332	&	-I	&	O-C Gateway	\\
56186.3665	&	2012.8	&	1360.5	&	0.00801	&	-I	&	O-C Gateway	\\
56483.4310	&	2013.6	&	1712.0	&	0.00169	&	-I	&	O-C Gateway	\\
56487.6559	&	2013.6	&	1717.0	&	0.00083	&	V	&	O-C Gateway	\\
56494.4181	&	2013.6	&	1725.0	&	0.00182	&	-I	&	O-C Gateway	\\
56496.5367	&	2013.7	&	1727.5	&	0.00754	&	-I	&	O-C Gateway	\\
56535.4043	&	2013.8	&	1773.5	&	-0.00184	&	-I	&	O-C Gateway	\\
56810.5034	&	2014.5	&	2099.0	&	0.00038	&	-I	&	O-C Gateway	\\
56924.5950	&	2014.8	&	2234.0	&	-0.0035	&	vis	&	O-C Gateway	\\
56924.5997	&	2014.8	&	2234.0	&	0.0012	&	V	&	O-C Gateway	\\
57198.4295	&	2015.6	&	2558.0	&	0.00185	&	-I	&	O-C Gateway	\\
57201.8101	&	2015.6	&	2562.0	&	0.00184	&	V	&	O-C Gateway	\\
57206.4575	&	2015.6	&	2567.5	&	0.00095	&	Clear	&	O-C Gateway	\\
57214.4880	&	2015.6	&	2577.0	&	0.00247	&	-I	&	O-C Gateway	\\
57219.5606	&	2015.6	&	2583.0	&	0.00416	&	-I	&	O-C Gateway	\\
57225.4742	&	2015.6	&	2590.0	&	0.0017	&	-I	&	O-C Gateway	\\
57260.5559	&	2015.7	&	2631.5	&	0.0096	&	-Ir	&	O-C Gateway	\\
57267.7314	&	2015.8	&	2640.0	&	0.00131	&	V	&	O-C Gateway	\\
57574.5243	&	2016.6	&	3003.0	&	0.00414	&	-I	&	O-C Gateway	\\
57657.3476	&	2016.8	&	3101.0	&	0.00254	&	Clear	&	O-C Gateway	\\
57924.4177	&	2017.6	&	3417.0	&	0.00474	&	-Ir	&	O-C Gateway	\\
57926.5294	&	2017.6	&	3419.5	&	0.00356	&	-Ir	&	O-C Gateway	\\
57995.4110	&	2017.8	&	3501.0	&	0.0053	&	CCD	&	O-C Gateway	\\
58265.8611	&	2018.5	&	3821.0	&	0.00685	&	V	&	O-C Gateway	\\
58306.4286	&	2018.6	&	3869.0	&	0.00711	&	R	&	O-C Gateway	\\
58327.5577	&	2018.7	&	3894.0	&	0.00738	&	-Ir	&	O-C Gateway	\\
58344.4606	&	2018.7	&	3914.0	&	0.00722	&	BVRI	&	O-C Gateway	\\
58414.6086	&	2018.9	&	3997.0	&	0.00766	&	V	&	O-C Gateway	\\
58749.2881	&	2019.8	&	4393.0	&	0.00711	&	V	&	O-C Gateway	\\
58749.2884	&	2019.8	&	4393.0	&	0.00733	&	V	&	O-C Gateway	\\
58749.2884	&	2019.8	&	4393.0	&	0.00737	&	V	&	O-C Gateway	\\
59035.7977	&	2020.6	&	4732.0	&	0.01026	&	V	&	O-C Gateway	\\
59083.1224	&	2020.7	&	4788.0	&	0.00646	&	V	&	O-C Gateway	\\
59337.5131	&	2021.4	&	5089.0	&	0.00647	&	V	&	O-C Gateway	\\
59337.5148	&	2021.4	&	5089.0	&	0.00815	&	V	&	O-C Gateway	\\
59381.4637	&	2021.5	&	5141.0	&	0.00921	&	-Ir	&	O-C Gateway	\\
59770.2357(3)	&	2022.6	&	5601.0	&	0.01140	&		&	{\it TESS}	\\
59770.6608(10)	&	2022.6	&	5601.5	&	0.01399	&		&	{\it TESS}	\\
59771.0807(2)	&	2022.6	&	5602.0	&	0.01125	&		&	{\it TESS}	\\
59771.5033(6)	&	2022.6	&	5602.5	&	0.01130	&		&	{\it TESS}	\\
59771.9257(2)	&	2022.6	&	5603.0	&	0.01113	&		&	{\it TESS}	\\
59772.3482(4)	&	2022.6	&	5603.5	&	0.01101	&		&	{\it TESS}	\\
59772.7705(2)	&	2022.6	&	5604.0	&	0.01076	&		&	{\it TESS}	\\
59773.1928(6)	&	2022.6	&	5604.5	&	0.01046	&		&	{\it TESS}	\\
59773.6174(3)	&	2022.6	&	5605.0	&	0.01247	&		&	{\it TESS}	\\
59774.0410(9)	&	2022.6	&	5605.5	&	0.01352	&		&	{\it TESS}	\\
59774.4619(2)	&	2022.6	&	5606.0	&	0.01188	&		&	{\it TESS}	\\
59774.8823(8)	&	2022.6	&	5606.5	&	0.00967	&		&	{\it TESS}	\\
59775.3012(21)	&	2022.6	&	5607.0	&	0.00603	&		&	{\it TESS}	\\
59775.7303(10)	&	2022.6	&	5607.5	&	0.01259	&		&	{\it TESS}	\\
59776.1535(15)	&	2022.6	&	5608.0	&	0.01313	&		&	{\it TESS}	\\
59776.5744(3)	&	2022.6	&	5608.5	&	0.01151	&		&	{\it TESS}	\\
59779.5318(2)	&	2022.6	&	5612.0	&	0.01083	&		&	{\it TESS}	\\
59779.9543(4)	&	2022.6	&	5612.5	&	0.01083	&		&	{\it TESS}	\\
59780.3776(2)	&	2022.6	&	5613.0	&	0.01152	&		&	{\it TESS}	\\
59780.8015(7)	&	2022.6	&	5613.5	&	0.01286	&		&	{\it TESS}	\\
59781.2233(3)	&	2022.6	&	5614.0	&	0.01206	&		&	{\it TESS}	\\
59781.6430(12)	&	2022.6	&	5614.5	&	0.00919	&		&	{\it TESS}	\\
59782.0673(3)	&	2022.6	&	5615.0	&	0.01087	&		&	{\it TESS}	\\
59783.7602(9)	&	2022.6	&	5617.0	&	0.01348	&		&	{\it TESS}	\\
59784.1754(7)	&	2022.6	&	5617.5	&	0.00613	&		&	{\it TESS}	\\
59784.6029(4)	&	2022.6	&	5618.0	&	0.01105	&		&	{\it TESS}	\\
59785.0251(2)	&	2022.6	&	5618.5	&	0.01072	&		&	{\it TESS}	\\
59785.4475	&	2022.7	&	5619.0	&	0.01054	&	V	&	O-C Gateway	\\
59785.4519(16)	&	2022.6	&	5619.0	&	0.01486	&		&	{\it TESS}	\\
59785.8711(3)	&	2022.7	&	5619.5	&	0.01151	&		&	{\it TESS}	\\
59786.2934(2)	&	2022.7	&	5620.0	&	0.01128	&		&	{\it TESS}	\\
59786.7153(10)	&	2022.7	&	5620.5	&	0.01061	&		&	{\it TESS}	\\
59787.1391(3)	&	2022.7	&	5621.0	&	0.01183	&		&	{\it TESS}	\\
59787.5603(6)	&	2022.7	&	5621.5	&	0.01047	&		&	{\it TESS}	\\
59787.9834(2)	&	2022.7	&	5622.0	&	0.01090	&		&	{\it TESS}	\\
59788.4073(10)	&	2022.7	&	5622.5	&	0.01225	&		&	{\it TESS}	\\
59788.8293(3)	&	2022.7	&	5623.0	&	0.01165	&		&	{\it TESS}	\\
59789.2493(8)	&	2022.7	&	5623.5	&	0.00916	&		&	{\it TESS}	\\
59789.6756(3)	&	2022.7	&	5624.0	&	0.01284	&		&	{\it TESS}	\\
59790.0968(1)	&	2022.7	&	5624.5	&	0.01147	&		&	{\it TESS}	\\
59790.5193(17)	&	2022.7	&	5625.0	&	0.01137	&		&	{\it TESS}	\\
59792.2097(4)	&	2022.7	&	5627.0	&	0.01145	&		&	{\it TESS}	\\
59792.6323(4)	&	2022.7	&	5627.5	&	0.01150	&		&	{\it TESS}	\\
59793.0543(1)	&	2022.7	&	5628.0	&	0.01092	&		&	{\it TESS}	\\
59793.4782(7)	&	2022.7	&	5628.5	&	0.01222	&		&	{\it TESS}	\\
59793.8999(2)	&	2022.7	&	5629.0	&	0.01135	&		&	{\it TESS}	\\
59794.3236(7)	&	2022.7	&	5629.5	&	0.01255	&		&	{\it TESS}	\\
59794.7446(3)	&	2022.7	&	5630.0	&	0.01091	&		&	{\it TESS}	\\
59795.1693(8)	&	2022.7	&	5630.5	&	0.01301	&		&	{\it TESS}	\\
59795.5905(2)	&	2022.7	&	5631.0	&	0.01164	&		&	{\it TESS}	\\
59796.0130(1)	&	2022.7	&	5631.5	&	0.01162	&		&	{\it TESS}	\\

59812.4930	&	2022.7	&	5651.0	&	0.01114	&	V	&	O-C Gateway	\\
59839.5376	&	2022.8	&	5683.0	&	0.01089	&	V	&	O-C Gateway	\\
60195.3471	&	2023.8	&	6104.0	&	0.01153	&	B	&	O-C Gateway	\\
60195.3472	&	2023.8	&	6104.0	&	0.01158	&	B	&	O-C Gateway	\\
60195.3472	&	2023.8	&	6104.0	&	0.01166	&	B	&	O-C Gateway	\\
60195.3475	&	2023.8	&	6104.0	&	0.01195	&	B	&	O-C Gateway	\\
60536.3619	&	2024.7	&	6507.5	&	0.00763	&	B	&	O-C Gateway	\\
60583.2694(2)	&	2024.8	&	6563.0	&	0.00919	&	V	&	In this study	\\

\end{longtable}

\end{document}